\documentclass[11pt]{article}
\usepackage{sao1}
\usepackage{graphicx}

\def\iue{\it IUE\/}

\def\units{$10^{-10}\mathrm{erg}\hspace{0.8mm}\mathrm{s}^{-1}\mathrm{cm}^{-2}\mathrm{\AA}^{-1}$}
\begin{document}
%
%
%
%
\title{Variations of Flux Intensity in Large Features and Spectral Lines of $\alpha^{\rm 2}$~CVn in Ultraviolet
\thanks{Based on $INES$ data from the $IUE$ satellite.}}
\author{N. A. Sokolov}
\institute{Central Astronomical Observatory at Pulkovo,
St. Petersburg 196140, Russia \\ e-mail:~sokolov@gao.spb.ru}
\maketitle
\begin{abstract}
Variation of intensity of the flux at the cores of large features and spectral lines of the classical magnetic CP star $\alpha^{\rm 2}$~CVn in the ultraviolet spectral region from 1150 to 3200~\AA\ is investigated.
This study is based on the archival {\it International Ultraviolet Explorer\/}
data obtained at different phases of the rotational cycle.
The shapes of two light curves at $\lambda\lambda$~1375 and 1415~\AA\ at the core
of large feature at $\lambda$~1400~\AA\ curves significantly differ.
The light curve at $\lambda$~1375~\AA\ show similar shape as in the 'pseudo-continuum'. The same behavior of the flux is at the cores of Si~II
resonance lines at $\lambda\lambda$~1260--64 and 1485~\AA.
While, the light curve at $\lambda$~1415~\AA\ show the phase displacement of
minimum of the flux. The phase displacement also is presented at the cores of
Si~II resonance lines at $\lambda\lambda$~1304--09 and 1485~\AA.
The minimal values of the amplitude of the flux variations are reached
at the cores of the large features at $\lambda\lambda$~1560 and 1770 \AA\ and at the cores of the strong Si~II resonance lines.
The flux at the cores of large feature at $\lambda$~1770~\AA\ and Fe~II resonance line at $\lambda$~1725--31~\AA, within errors measurements, does not vary.
The investigation of variability of the flux in the wings of ${\rm Ly}_{\alpha}$
line indicate that the fluxes, which formed in inner layers of atmosphere, are redistributed into outer layers of atmosphere of $\alpha^{\rm 2}$~CVn.


\end{abstract}

\section {Introduction}
The magnetic Chemically Peculiar (mCP) star $\alpha^{\rm 2}$~CVn (HD~112413,
HR~4915) displays strong line profile variations in the visual spectral region, attributable to the non-uniform chemical abundance distribution on stellar surface, particularly of the lines of Fe, Cr and Ti (Khokhlova \& Pavlova 1984), Eu, Cr and Si (Goncharskii et al. 1983) and O (Gonzalez \& Artru 1994). Ryabchikova et al. (1999) report the first identification of the Eu~III $\lambda$~6666.347 line in spectrum $\alpha^{\rm 2}$~CVn. Later Kochukhov et al. (2002) used the new magnetic Doppler Imaging code to reconstruct the magnetic field geometry and surface abundance distributions for six chemical elements of this star.

The study in the ultraviolet have been restricted to much lower resolution, usually without resolving individual stellar lines (e.g., Leckrone \& Snijders~1979).
Hensberge et al. (1986) identified the ions Mg~II Si~II, Cr~II, Mn~II, Fe~II, Fe~III, Co~II, Ni~II, Yb~II and W~II in the mid-ultraviolet spectrum of $\alpha^{\rm 2}$~CVn, which has been obtained with the BUSS (Balloon-Borne Ultraviolet Spectrograph).
High-dispersion IUE observations of the two resonance lines of Mg~II, at $\lambda\lambda$ 1650 and 1942 \AA, were  investigated by Leckrone (1984) for
two normal stars, six HgMn stars and for the magnetic variable $\alpha^{\rm 2}$~CVn.
The last star possesses moderately strong Hg~II resonance lines, confirming its classification as Hg-rich, but its magnesium anomaly is less pronounced than that
of the MgMn stars. Fuhrmann~(1989) investigated the high-resolution spectra of the CP star HR 465. For comparison purposes, the spectra of some other stars, including $\alpha^{\rm 2}$~CVn, were discussed as well. The author have noted that the C~II resonance doublet at $\lambda$~1334.5--1335.7~\AA\ are comparatively weak in the spectrum of $\alpha^{\rm 2}$~CVn.

To quantify the ultraviolet variations of the C~IV doublet at $\lambda\lambda$~1548, 1550~\AA, Shore et al.~(1987) have formed the photometric line index, expressed in magnitudes. Later, Sokolov~(2000, 2006) had introduced the analogous photometric indices in order to derive the variations of  the total absorption in the broad features at far-UV spectral region.
Unfortunately, these indices depends strongly upon the stability of the intensity
of the near continuum. Based on the fact that the {\iue} Newly Extracted Spectra (INES) data are presented in absolute units, it is possible to investigate the variations of the fluxes at the cores of the large features and spectral lines.
Recently, such attempt was made by Sokolov~(2010) for mCP star 56~Ari.
Another mCP star is $\alpha^{\rm 2}$~CVn for which there are enough {\iue} data in order to investigate the variability of the flux at the cores of the large features and spectral lines. This is done in the present paper.

\section{Observational Data and Analysis}

\subsection{{\iue} spectra}

The {\iue} spectra used in this study are low-resolution echelle spectra
obtained with a resolution of about 6~\AA. Additionally, the 'rebinned' spectra
from high-dispersion images of $\alpha^{\rm 2}$~CVn were used, as well.
In all cases, the spectra were obtained through the large aperture (9.5$\arcsec$~$\times$~22$\arcsec$). Finally, we analyzed 22 SWP, 10 LWR and 6 LWP spectra, distributed quite smoothly over the period of rotation. The description of these {\iue} spectra in detail is done  by Sokolov~(2011).

\subsection{Data analysis}

To analyze the {\iue} spectra of $\alpha^{\rm 2}$~CVn we used the linearized least-squares method. An attempt was made to describe the light curves in a
quantitative way by adjusting a Fourier series.
The method has already applied to the {\iue} data of $\alpha^{\rm 2}$~CVn
and has shown the very good descriptions of the monochromatic light curves in the 'pseudo-continuum' (Sokolov~2011).
Thus, the observations were also fitted by a simple cosine wave:
\begin{equation}
F(\lambda,\,t)=A_{0}(\lambda) +
A_{1}(\lambda)\cos(2\pi(t-T_{0})/P +\phi(\lambda)),
\label{equation_2}
\end{equation}
where $F(\lambda,\,t)$ is a flux for the given $\lambda$ and the $t$ is Julian date of the observation.
The $T_{0}$ and $P$ are zero epoch and rotational period of Farnsworth~(1932) ephemeris, respectively.
The coefficients $A_{0}(\lambda)$ of the fitted curves
define the average distribution of energy over the cycle of the variability while
the coefficients $A_{1}(\lambda)$ define the semi-amplitude of the flux variations
for the given $\lambda$.

\section{Identification of Large Features and Spectral Lines in the Spectrum of $\alpha^{\rm 2}$~CVn} \label{sect_3}

In the far-UV spectral region silicon appears as the main absorber with
the strong resonance lines at $\lambda\lambda$~1260--64, 1304--09, 1485, 1526--33~\AA\ (see Fig.~2 of Sokolov~2011).
It should be noted that the blend at $\lambda$1304--09~\AA\ has two major contributors: the resonance doublet and the autoionising multiplets
(Artru \& Lanz~1987).
According to Artru \& Lanz~(1987), the strong lines in the spectrum
of CP~stars appear from C~II at $\lambda$~1334~\AA\ and Al~II at $\lambda$~1671~\AA, which is close to a strong C~I at $\lambda$~1657~\AA\ line. Fe~II form a depression at $\lambda$~1725--31~\AA\ and Cr~II produces line at $\lambda$~1434~\AA. Although in the case of $\alpha^{\rm 2}$~CVn the line from
C~II at $\lambda$~1334~\AA\ is not detectable at this resolution.
Moreover, the C~II resonance doublet at $\lambda$~1334.5--1335.7~\AA\ are comparatively weak at the high-resolution spectrum of this star (Fuhrmann~1989).

\begin{figure*}
\centerline{\includegraphics[width=150mm, angle=0]{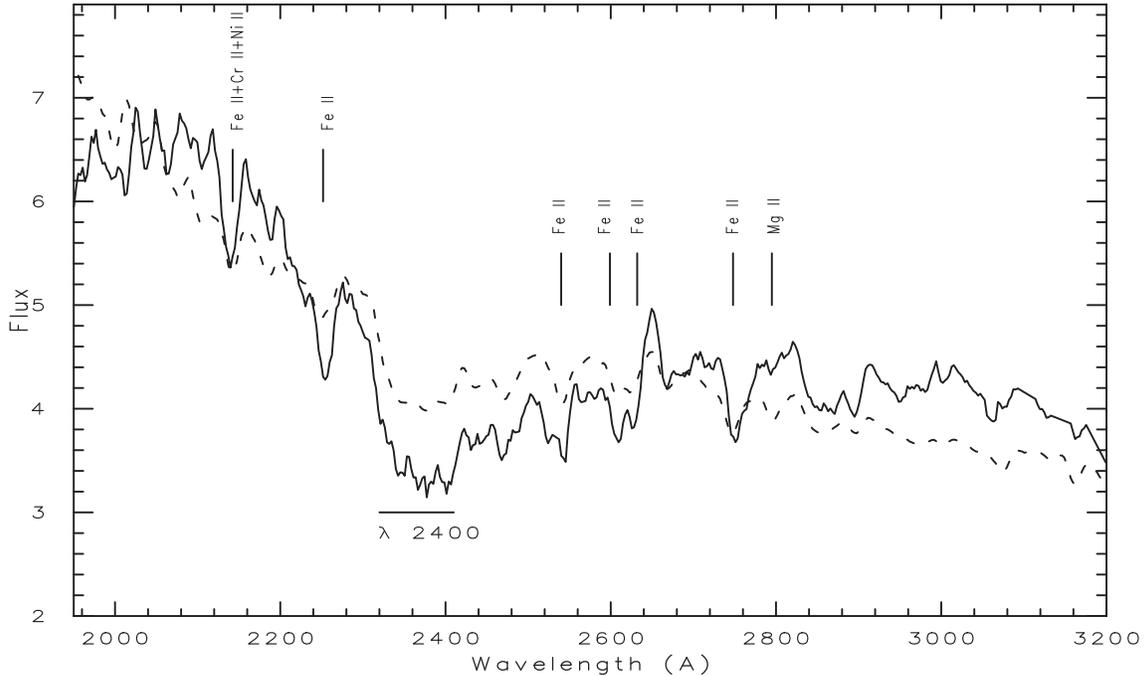}}
\caption{The average distributions of energy in {\units} of $\alpha^{\rm 2}$~CVn
 (solid line) and 56~Ari (dashed line).
 The prominent spectral lines and features are shown by vertical and
 horizontal lines, respectively. (see text)}
\label{mean_lw}
\end{figure*}

In previous study of the star 56~Ari we identified which elements are
responsible for depressions of the flux centered at $\lambda\lambda$~2140, 2250,
2540, 2607, 2624 and 2747~\AA\ (Sokolov~2010).
The synthetic spectrum calculation showed that Fe~II appears as the
main absorber for these depressions. On the other hand, the same
synthetic spectrum calculation showed that mainly Fe, Cr and Ni are responsible
for depression of the flux at $\lambda$~2140~\AA. Comparison of the average energy distribution of $\alpha^{\rm 2}$~CVn with the average energy distribution
of 56~Ari showed that the depressions of the flux centered at $\lambda\lambda$~2140, 2250, 2540, 2607, 2624 and 2747~\AA\ also is presented in the spectrum of
$\alpha^{\rm 2}$~CVn. It is well known that Mg~II resonance lines at $\lambda\lambda$~2795, 2798 and 2803~\AA\ are responsible for depression of
the flux at $\lambda$~2800~\AA. This depression is not detectable at low-resolution
mode in the spectrum of $\alpha^{\rm 2}$~CVn. Prominent depressions of the flux in the near-UV spectral region are indicated in Fig.~\ref{mean_lw} with their identification.
In order to compare the average distributions of energy of  $\alpha^{\rm 2}$~CVn
and 56~Ari, the fluxes of 56~Ari were increased by the factor ten on Fig.~\ref{mean_lw}.

Four large features at $\lambda\lambda$~1400, 1560, 1770 and 2350--2400~\AA,
which are strongly enhanced in the spectrum of CP stars, are well seen in
spectrum of $\alpha^{\rm 2}$~CVn. Lanz et al.~(1996) have given strong
arguments supporting the idea that the intense autoionization resonanses
of Si~II could explain the features at $\lambda$~1400 and 1560~\AA\ in
the spectrum of CP~stars.
On the other hand, they were unable to identify the depression of the flux
at $\lambda$~1770~\AA. Another element may cause this strong depression.
Comparison of the {\iue} high resolution spectrum
of 56~Ari with the full synthetic spectrum as well as those including
lines from only one element showed that the iron is responsible for
depression of the flux at $\lambda$~1770~\AA\ (see Fig.~9 of Sokolov~2006).
It is necessary to note that  this depression  is considerably increased in spectrum
of $\alpha^{\rm 2}$~CVn than in spectrum of 56~Ari.
The large feature at $\lambda$~2350--2400~\AA\, which are strongly
enhanced in the spectrum of CP stars, are well seen in spectrum of
$\alpha^{\rm 2}$~CVn. The lines of iron peak elements have a particularly
important contribution to opacity at $\lambda$~2350--2400~\AA.
Adelman et al.~(1993) have given strong arguments supporting the idea that
a large number the lines of iron can explain the feature at $\lambda$~2350--2400~\AA.
Many CP~stars have here a very pronounced depression of the flux compared to
normal stars (Stepie{\' n} \& Czechowski~1993).

\section{Average Flux Determination at the Cores of Large Features and
Spectral Lines in the Spectrum of $\alpha^{\rm 2}$~CVn}

To measure the absorption at the cores of large features, the spectra were processed using the spectral reduction software {\sc spe} developed by S. Sergeev at the Crimean Astrophysical Observatory (CrAO).
The program allows measuring the average intensity of the flux and corresponding error in any selected rectangular spectral region.
For the large features at $\lambda$~1560 and 1770~\AA\ spectral regions are $\sim$10~\AA\ wide, while for the largest feature at  $\lambda$~1400~\AA\ two spectral regions were selected and centered at $\lambda$~1375 and 1415~\AA. On the other hand, for the large feature at $\lambda$~2350--2400~\AA\ the spectral region was used with $\sim$50~\AA\ wide and centered at $\lambda$~2375~\AA.

\begin{figure}[t]
\vspace{-0.2cm}
\centerline{\includegraphics[width=85mm]{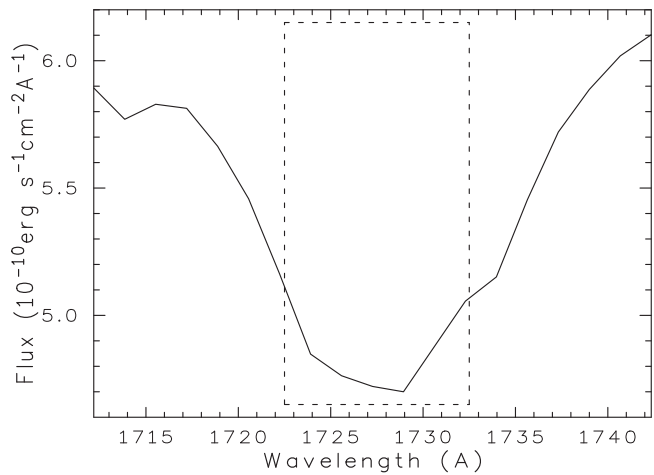}}
\caption{The spectral region of Fe II doublet at $\lambda$~1725-31 \AA\
 of $\alpha^{\rm 2}$~CVn. The rectangle shows the selected spectral region for calculation of the average intensity at the core of this doublet.}
\label{a0_1725}
\end{figure}

The resonance doublets of Si~II and Fe~II lines are looked as depressions in the {\iue} low-resolution spectrum of $\alpha^{\rm 2}$~CVn, as illustrated by Fig.~\ref{a0_1725}.
Thus, the average intensity of the flux at the cores of the spectral lines was computed by averaging five nearest fluxes for a given $\lambda$:
\begin{equation}
 F(\lambda) = \frac{1}{5}\sum_{i=1}^{5} F(\lambda-\lambda_{step\cdot(i-3)})
\label{equation_2}
\end{equation}
where $\lambda_{step}$ is equal 1.676~\AA\ for SWP camera and is equal 2.669~\AA\
for LWR and LWP cameras.
As far as the errors in $F(\lambda)$ are concerned, we computed them by taking
into account the the errors in the fluxes as presented in INES $Catalog$,
according to the standard propagation theory of errors.
In order to check reliability of the average intensity of the flux at the cores
of the spectral lines, the spectra were also processed using program {\sc spe}.
Experience showed that within errors of measurements the average intensity of the fluxes computed by program {\sc spe} and by using equation~\ref{equation_2}
are the same.

\subsection{Variations of Large Features}

\begin{figure*}[tp]
\vspace{-0.2cm}
\centering \resizebox{0.48\hsize}{!}{\includegraphics{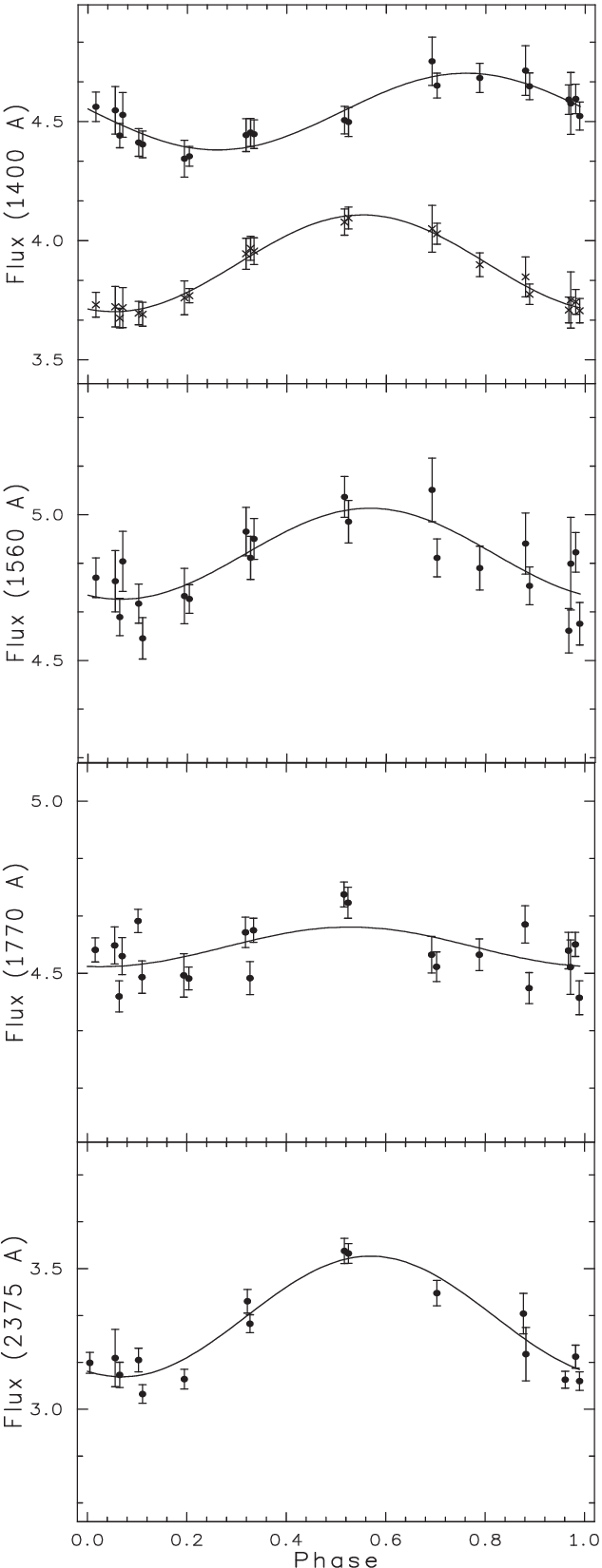}}
\hspace{+0.4cm}
\centering \resizebox{0.48\hsize}{!}{\includegraphics{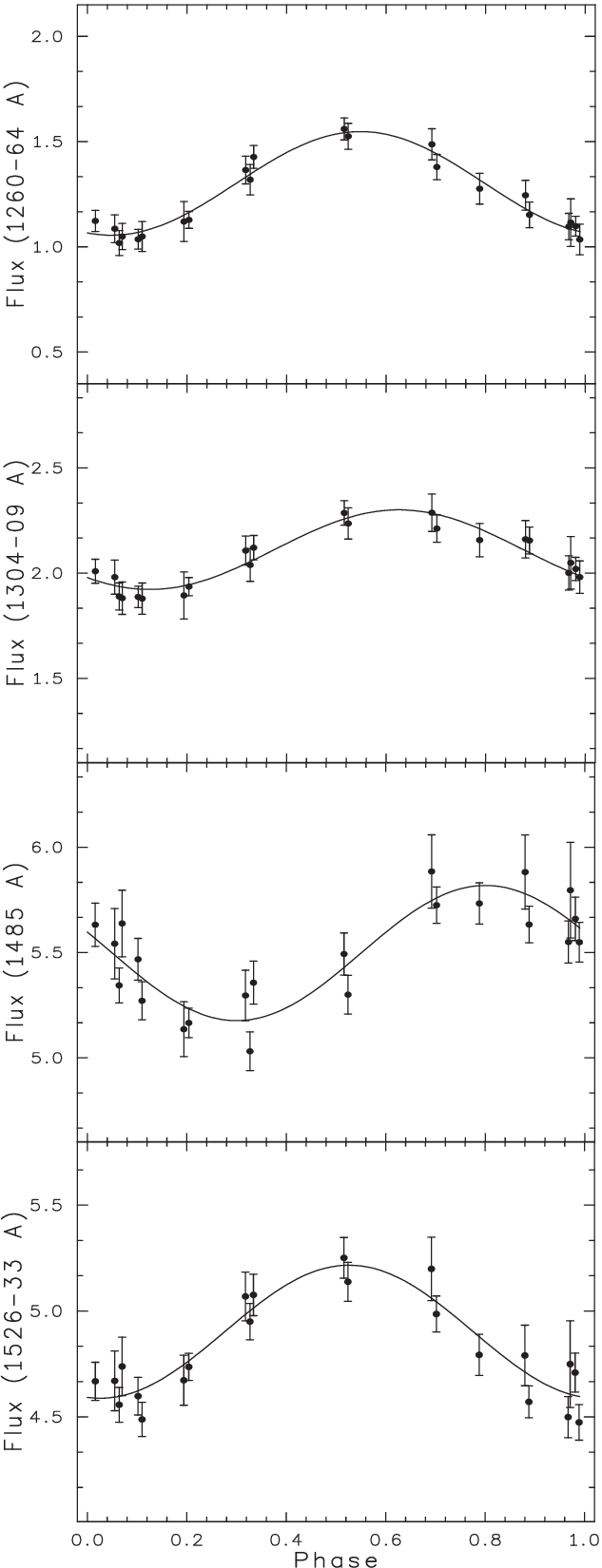}}
\caption{{\bf Left:} Phase diagrams of the light curves at the cores of large features in {\units} for $\alpha^{\rm 2}$~CVn.
The upper panel represent two phase diagrams centered at $\lambda$~1375~\AA\ (crosses) and at $\lambda$~1415~\AA\ (circles).
1$\sigma$ error bars accompany each data point.
Solid lines denote the fit according to equation~\ref{equation_2}.
{\bf Right:} Phase diagrams of the light curves in at the cores of Si II resonance lines.}
\label{features}
\end{figure*}

Figure~\ref{features} exhibits the variations of the average intensity of
the fluxes at the cores of large features versus the rotational phase.
First of all, the shapes of two light curves at the core of large feature
at $\lambda$~1400~\AA\ significantly differ.
The minimum of the light curve at $\lambda$~1375~\AA\ is reached at phase 0.05
while the minimum of the light curve at $\lambda$~1415~\AA\ is at phase 0.26.
Although, the amplitudes of two light curves at $\lambda\lambda$~1375 and 1415~\AA\ are approximately the same and are equal to 0.41 and 0.32~$\times$~{\units}, respectively.
According to the model computation of Lanz et al.~(1996), two features at $\lambda\lambda$~1400 and 1560~\AA\ are connected with the intense autoionization resonance of Si~II.
Qualitative comparison of the light curves at $\lambda\lambda$~1375 and 1560~\AA\
shows the good agreement.
The minimum of the light curve at $\lambda$~1375~\AA\ is reached at phase 0.05
while the minimum of the light curve at $\lambda$~1560~\AA\ is at phase 0.10.
Also, the amplitudes of two light curves at $\lambda\lambda$~1375 and 1560~\AA\
are approximately the same and are equal to 0.41 and 0.38~$\times$~{\units},
as illustrated by Fig.~\ref{features}.

The identification of two features at $\lambda\lambda$~1770 and 2375~\AA\
are connected with a large concentration of iron lines (see Sect.~\ref{sect_3}).
The flux variations at the cores of the features at $\lambda\lambda$~1770 and
2375~\AA\ also significantly differ.
Thus, the amplitude of the light curve at the core of the feature at $\lambda$~1770~\AA\ is equal to 0.11~$\times$~{\units}.
Practically, the flux at the core of this feature, within errors  measurements,
does not vary. On the other hand, the amplitude of the light curve at the core
of the feature at $\lambda$~2375~\AA\ is equal to 0.43~$\times$~{\units}.
Although, the shapes of light curves are approximately the same, as illustrated by Fig.~\ref{features}. One can see, that the behavior of the flux is different at the cores of large features, even if the same element is responsible for the features.
This puzzling situation was one of the incentives for studying the variations
of the flux at the cores of the Si~II resonance lines and the depressions for
which is responsible a concentration of Fe~II lines.

\subsection{Variations of Si~II Resonance Lines}

Even for the normal stars, Si~II appears as the main absorber with
the strongest resonance lines recognizable at $\lambda\lambda$~1260--64,
1304--09, 1485, 1526--33~\AA\ (Artru \& Lanz 1987).
Figure~\ref{features} exhibits the variations of the flux
at the cores of these depressions versus the rotational phase.
Note that the vertical scales are the same for each part of the figure.
One can see from Fig.~\ref{features}, the light curves have the similar shapes
in cores of resonance lines at $\lambda\lambda$~1260--64 and 1526--33~\AA.
The minima of the light curves at $\lambda\lambda$~1260--64
and 1526--33~\AA\ are reached at phases 0.05 and 0.02, respectively.
The similar behavior also shows the nearest monochromatic light curves in the 'pseudo-continuum' (see Fig.~3 of Sokolov~2011). Also, the amplitudes of
the light curves at the cores of the depressions at $\lambda\lambda$~1260--64
and 1526--33~\AA\ are in the good agreement and are equal to 0.49 and 0.63~$\times$~{\units}, respectively. It should be noted that the similar shape
shows the light curve at $\lambda$~1375~\AA.
On the other hand, the minima of the light curves at $\lambda\lambda$~1304--09 and 1485~\AA\ are reached at phases 0.12 and 0.30, as illustrated by Fig.~\ref{features}.
Although, the minima of the monochromatic light curves are reached at phase
$\sim$0.0 in the nearest 'pseudo-continuum'.
Moreover, the amplitudes of the light curves
are different at $\lambda\lambda$~1304--09 and 1485~\AA\ and are equal 0.38 and 0.64~$\times$~{\units}, respectively. It should be noted that the similar shape
shows the light curve at $\lambda$~1415~\AA. The monochromatic light curves in
the near-UV spectral region with $\lambda$~$>$~2505~\AA\ in the 'pseudo-continuum'
also shows such behavior.
The near-UV spectral region is quite important in order to investigate
the variability of the flux at the cores of Fe~II lines.

\begin{figure*}[t]
\vspace{-0.2cm}
\centerline{\includegraphics[width=170mm, angle=0]{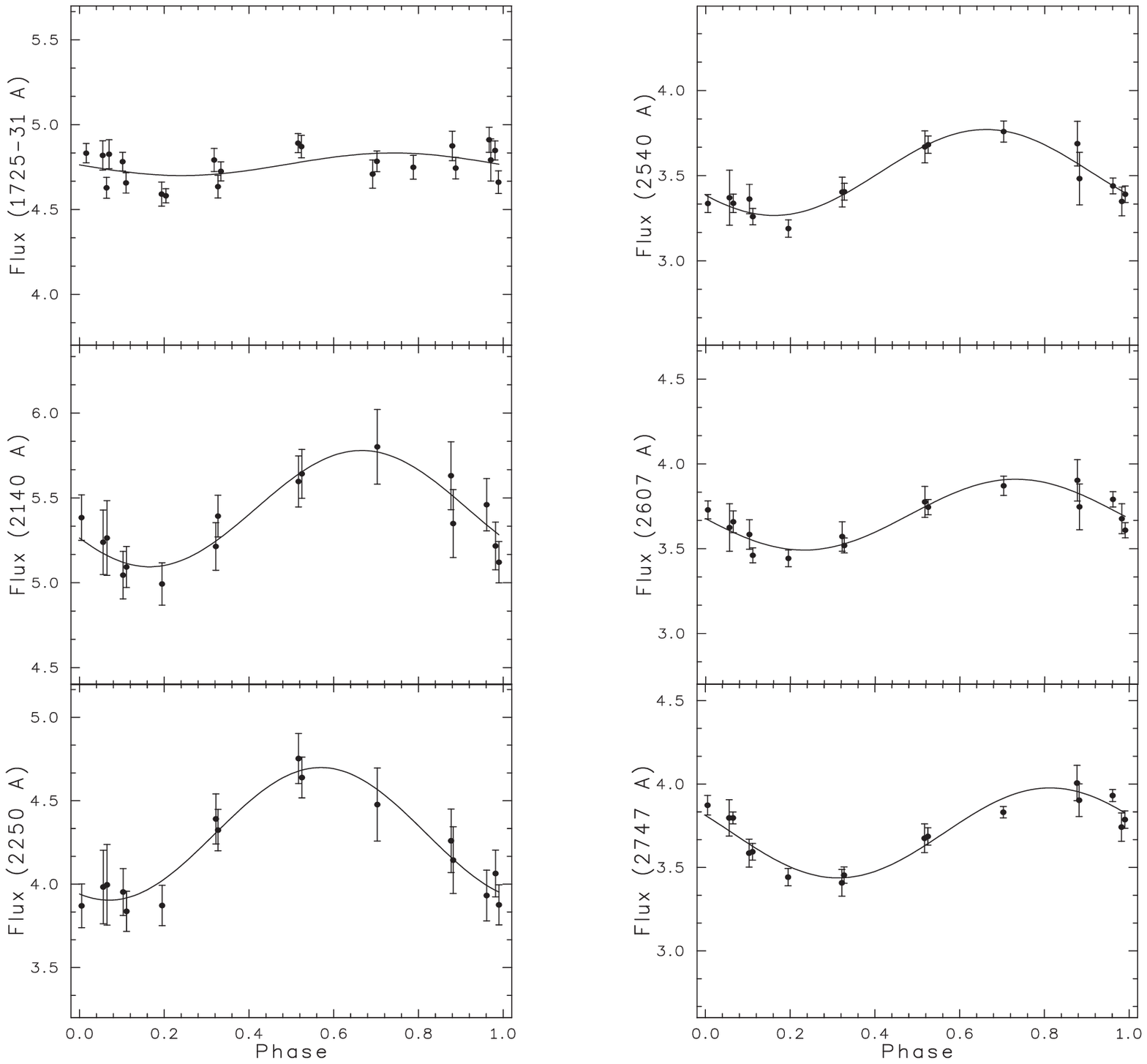}}
\caption{The same as in Fig.~\ref{features} for the light curves at the cores
 of Fe~II lines.}
\label{iron}
\end{figure*}

\subsection{Variations of Fe~II Lines}

In the far-UV spectral region Fe~II doublet at $\lambda$~1725--31~\AA\ gives only
one depression of the flux (Artru \& Lanz~1987).
The synthetic spectrum calculation showed that Fe~II appears as the
main absorber of the flux at $\lambda\lambda$~2250, 2540, 2607, 2624 and 2747~\AA\
(see Sect.~\ref{sect_3}). Additionally, the depression of the flux at $\lambda$~2140~\AA\ was included in our investigation. On the other hand, two depressions of the flux at $\lambda\lambda$~2607 and 2624~\AA\ showed the same behavior. Thus, we have included in our investigation only the depression of the flux at $\lambda$~2607~\AA. Figure~\ref{iron} exhibits the variations of the average intensity of the flux at the cores of Fe~II lines versus the rotational phase.
Note that the vertical scales are the same for each part of the figure.
As can be seen on the graphs of Fig.~\ref{iron}, the variability of the flux at the cores of Fe~II lines shows by about the same behavior as the monochromatic light curves with $\lambda$~$>$~2505~\AA\ in the 'pseudo-continuum' (see. Fig.~3 of Sokolov~2011). The light curves at the cores of of Fe~II lines are showing
the phase displacement of the minimum of the flux from 0.07 at $\lambda$~2250~\AA\
to 0.31 at $\lambda$~2747~\AA.
Additionally, the minimum of the light curve at the core of of Fe~II lines at $\lambda$~2140~\AA\ is at phase 0.16. Although, the amplitude of the light curve
at $\lambda$~2140~\AA\ is big enough and is equal to 0.69~$\times$~{\units}.
Possibly, it is because mainly three chemical elements (Fe, Cr, Ni) are responsible for this depression. The monochromatic light curves in the 'pseudo-continuum' only with $\lambda$~$>$~2505~\AA\ shows the same phase displacement of the minimum of the flux. It should be noted that the amplitudes of the light curves here are bigger than at the cores of Fe~II lines. On the other hand, the light curve at the core of Fe~II resonance line at $\lambda$~1725--31~\AA\ where the amplitude is equal to 0.13~$\times$~{\units}. As at the core of large feature at $\lambda$~1770~\AA, the flux does not vary at the core of this line.
Although, the variability of the flux in the nearest 'pseudo-continuum' is significant. Thus, the amplitudes of the monochromatic light curves at $\lambda\lambda$~1690 and 1794~\AA\ are equal 0.41 and 0.51~$\times$~{\units}, respectively.

\begin{figure}[t]
\vspace{-0.2cm}
\centerline{\includegraphics[width=95mm]{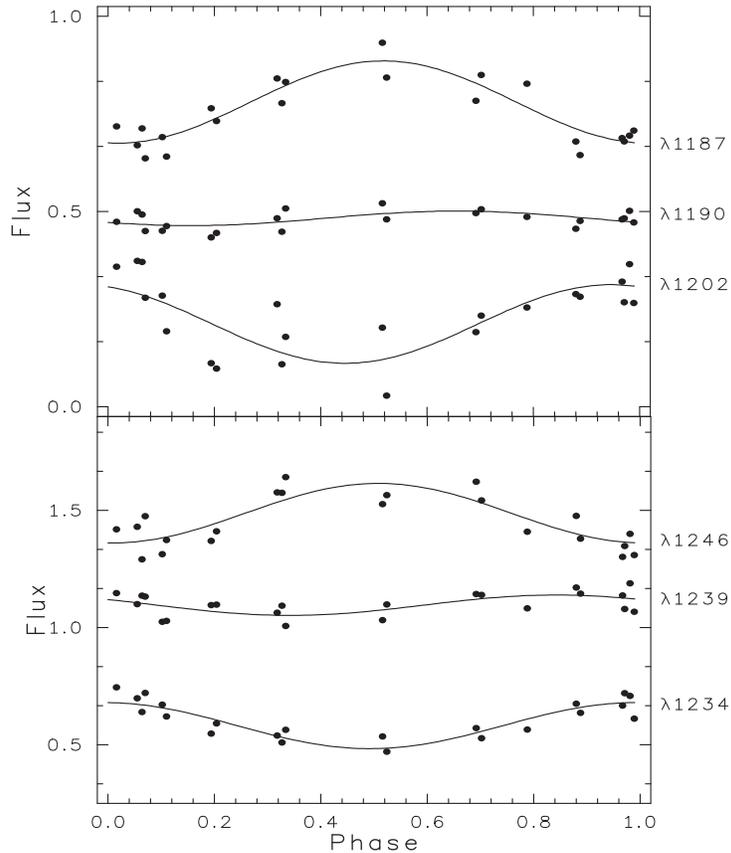}}
\caption{Phase diagrams of the light curves in {\units} in the wings of
 ${\rm Ly}_{\alpha}$ line for $\alpha^{\rm 2}$~CVn.
 Upper and lower panels show short-wavelength and long-wavelength sides of
 ${\rm Ly}_{\alpha}$ line, respectively.
 Solid lines denote the fit according to equation~\ref{equation_2}.}
\label{laim_com}
\end{figure}

\subsection{Variations of Lyman-Alpha Line}

Leckrone \& Snijders~(1979) have compared Lyman-alpha (${\rm Ly}_{\alpha}$)
profiles of $\alpha^{\rm 2}$~CVn at two phases 0.0 and 0.5, using $Copernicus$
data. The authors have drawn the conclusion that brightness variations at the core
of ${\rm Ly}_{\alpha}$ line are anomalous with respect to the adjacent ultraviolet regions. In addition, they noted that coverage of the complete cycle by a future space instrument will be necessary to establish the specific phasing of these variations.
Based on the fact that the star is enough bright ($m_{\rm v}$~=~2.90), the {\iue} data have allowed to investigate behavior of the  monochromatic light curves in short-wavelength and long-wavelength sides from the core of ${\rm Ly}_{\alpha}$ line. Several monochromatic light curves in the wings of ${\rm Ly}_{\alpha}$ line
at different wavelengths were formed.
Figure~\ref{laim_com} exhibits the variations of the flux in short-wavelength and long-wavelength sides of ${\rm Ly}_{\alpha}$ line versus the rotational phase.
As can be seen on the graphs of Fig.~\ref{laim_com} the monochromatic light curves have identical shapes in short-wavelength and long-wavelength sides of ${\rm Ly}_{\alpha}$ line.
The light curves at $\lambda\lambda$~1187 and 1246~\AA\ show the minimum and
maximum of the flux at phases 0.0 and 0.5, respectively.
Note that the monochromatic light curves in the 'pseudo-continuum' show the same behavior in the far-UV spectral region.
These monochromatic light curves are most removed from the line center of
${\rm Ly}_{\alpha}$ at $\lambda$~1215~\AA.
On the other hand, the monochromatic light curves  at $\lambda\lambda$~1202 and 1234~\AA\ show the minimum and maximum of the flux at phases 0.5 and 0.0, respectively. Also, the light curve in the $V$ filter shows the same behavior
in the visual spectral region (Pyper~1969).
While, the variation of the flux at $\lambda\lambda$~1190 and 1239~\AA\ is practically zero over the period of rotation. In other words, the so-called
'null wavelength regions' is also disposed in these wavelengths.
It should be noted that the fluxes at the cores of ${\rm Ly}_{\alpha}$
line varies with the small amplitudes at $\lambda$~1213~\AA\ for two CP~stars CU~Vir and 56~Ari (Sokolov~2000, 2006).
\begin{figure}[t]
\vspace{-0.2cm}
\centerline{\includegraphics[width=100mm]{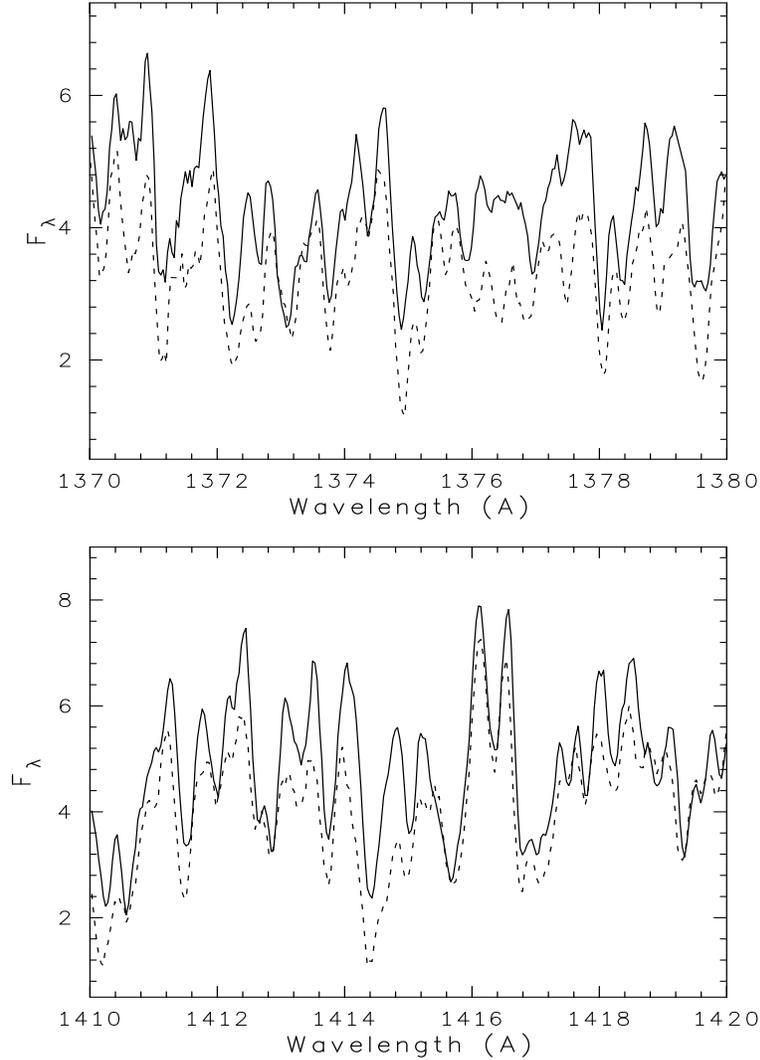}}
\caption{Two spectral regions of the broad feature at $\lambda$~1400~\AA\
 for $\alpha^{\rm 2}$~CVn. The high-dispersion spectra in the region of $\lambda$~1375~\AA~(top panel) and the high-dispersion spectra in the region
 of $\lambda$~1415~\AA~(bottom panel) (see the text).}
\label{1400n}
\end{figure}

\section{Discussion}

Our result indicate that the variations of average intensity of the flux at the cores of large features and spectral lines show different behavior.
First of all, the shapes of two light curves at $\lambda\lambda$~1375 and 1415~\AA\ at the core of large feature at $\lambda$~1400~\AA\ curves significantly differ.
The minimum of the light curve at $\lambda$~1375~\AA\ is reached at phase 0.05
while the minimum of the light curve at $\lambda$~1415~\AA\ is at phase 0.26.
Although, the difference in wavelength  between these spectral regions is equal 40~\AA.
This puzzling situation was one of the incentives for studying the high-dispersion spectra of $\alpha^{\rm 2}$~CVn in these spectral regions.
In the top panel of Fig.~\ref{1400n} two high-dispersion spectra are plotted in the spectral region near $\lambda$~1375~\AA.
These spectra SWP27894 and SWP15828 were obtained at the phases 0.064
(dashed line) and 0.524 (full line), respectively.
From Fig.~\ref{1400n} we see that the fluxes of the spectrum SWP27894
systematically lower than the fluxes of the spectrum SWP15828.
In the bottom panel of Fig.~\ref{1400n} two high-dispersion spectra are plotted in the spectral region near $\lambda$~1415~\AA.
These spectra SWP04813 and SWP27880 were obtained at the phases 0.204 (dashed line) and 0.702 (full line), respectively. Again, we can see that the fluxes of
the spectrum SWP04813 lower than the fluxes of the spectrum SWP27880.
The comparisons of the high-dispersion spectra in the spectral regions of
the broad feature at $\lambda$~1400~\AA\ are in agreement with phase diagrams obtained from the low-dispersion spectra (see Fig.~\ref{features}).
Although, the nature of such behavior of the fluxes in the spectral regions at $\lambda\lambda$ 1375 and 1415~\AA\ still unclear. Possibly, the influence of different species of spectral lines can play a some role.

The light curves at the cores of Si~II resonance lines at $\lambda\lambda$~1260--64
and 1526--33~\AA\ show the similar shapes as in the 'pseudo-continuum'.
While, the light curves at the cores of Si~II resonance lines at $\lambda\lambda$~1304--09 and 1485~\AA\ show the phase displacement
of minimum of the flux.
On the other hand, the flux at the cores of large feature at $\lambda$~1770~\AA\
and Fe~II resonance line at $\lambda$~1725--31~\AA, within errors measurements,
does not vary. Moreover, the light curves at the cores of of Fe~II lines show
the phase displacement of minimum of the flux from 0.07 at $\lambda$~2250~\AA\
to 0.31 at $\lambda$~2747~\AA.

The vertically dependent abundance stratification in CP stars are suggested by many authors (e.g., Ryabchikova~2008, and references therein).
This effect may influence our results, because the effective depth at which the fluxes is formed at $\lambda$~$<$~2000~\AA\ can differs from the effective depth at which the fluxes is formed at $\lambda$~$>$~2505~\AA.
We can expect that the flux comes from upper layers of atmosphere
at the cores of some large features and depressions than the flux coming from
the nearest 'pseudo-continuum' in the spectral region with $\lambda$~$<$~2000~\AA.
Therefore, possibly, some large features and depressions show the displacement
of position of minimum of the flux, though the flux in the nearest
'pseudo-continuum' does not show such displacement.
Although, the spectrum in the 'pseudo-continuum' is also blocked by a great number
of spectral lines of various chemical elements.

The main thing is that the energy blocking by silicon bound-free transitions and iron bound-bound transitions decreases the flux in the UV spectral region. The blocked flux appears in the visual and the red parts of the spectrum. Such an explanation is supported by the anti-phase relationship of light curves in the visual and the UV spectral regions.
Probably, our investigation indicates that not only this mechanism may influence on the redistribution of the flux in atmosphere of $\alpha^{\rm 2}$~CVn.
The fluxes, situated in the wings of ${\rm Ly}_{\alpha}$ line at different distances
from its center, and hence formed at different depths in stellar atmosphere.
The investigation of variability of the flux in the wings of ${\rm Ly}_{\alpha}$
line indicate that the fluxes, which formed in inner layers of atmosphere, are redistributed into outer layers of atmosphere of $\alpha^{\rm 2}$~CVn. Therefore, it should be some layer in atmosphere of the star where the fluxes do not vary over the period of rotation.
Although, the independent investigation of ${\rm Ly}_{\alpha}$ line variations for others CP~stars is needed in order to confirm our result.

\section {Conclusions}

The archival {\iue} spectrophotometric data of $\alpha^{\rm 2}$~CVn
have permitted to analyze the light variations at the cores of large features
and spectral lines. The variations of intensity of the flux at the cores of large features and spectral lines show different behavior.
The influence of different species of spectral lines play a some role
in the different regions at $\lambda\lambda$~1375 and 1415~\AA\ of the large
feature $\lambda$~1400~\AA.
The light curve at $\lambda$~1375~\AA\ show similar shape as in the 'pseudo-continuum'. The same behavior of the flux is at the cores of Si~II resonance lines at $\lambda\lambda$~1260--64 and 1485~\AA.
While, the light curve at $\lambda$~1415~\AA\ show the phase displacement of minimum
of the flux. The phase displacement also is presented at the cores of Si~II resonance lines at $\lambda\lambda$~1304--09 and 1485~\AA.
The same phase displacement is at the cores of the Fe~II depressions in the spectral region with $\lambda$~$>$~2505~\AA. But, the variability of the flux at the cores of the Fe~II depressions in this spectral region is the same as in
the 'pseudo-continuum'.

The minimal values of the amplitude of the flux variations are reached
at the cores of the large features at $\lambda\lambda$~1560 and 1770 \AA\ and at the cores of the strong Si~II resonance lines at $\lambda\lambda$~1260--64, 1304--09 and 1485~\AA. The flux at the cores of large feature at $\lambda$~1770~\AA\
and Fe~II resonance line at $\lambda$~1725--31~\AA, within errors measurements,
does not vary.

The investigation of variability of the flux in the wings of ${\rm Ly}_{\alpha}$
line indicate that the fluxes, which formed in inner layers of atmosphere, are redistributed into outer layers of atmosphere of $\alpha^{\rm 2}$~CVn.
Although, the independent investigation of ${\rm Ly}_{\alpha}$ line variations
in the high-resolution mode is needed in order to confirm our result.


\begin{thebibliography} {References}
\bibitem{} S.J. Adelman, C.R. Cowley, D.S. Leckrone, S.W. Roby, G.M. Wahlgren,
    1993, \apj, 419, 276,
\bibitem{} M.-C. Artru, T. Lanz, 1987, \aaa, 182, 273,
\bibitem{} G. Farnsworth, 1932, \apj, 75, 364,
\bibitem{} K. Fuhrmann, 1989, \aas, 80, 399,
\bibitem{} A.V. Goncharskii, T.A. Ryabchikova, V.V. Stepanov, V.L. Khokholova,
    A.G. Yagola, 1983, Soviet Astr., 14, 652,
\bibitem{} J.F. Gonzalez, M.-C. Artru, 1994, \aaa, 289, 209,
\bibitem{} H. Hensberge, J. Van Santvoort, K.A. Van der Hucht, T.H. Morgan,
    1986, \aaa, 158, 113,
\bibitem{} V.L. Khokhlova, V.M. Pavlova, 1984, PAZh, 10, 337,
\bibitem{} O. Kochukhov, N. Piskunov, I. Ilyin, I. Tuominen, 2002, \aaa,
    389, 420,
\bibitem{} T. Lanz, M.-C. Artru, M. Le Dourneuf, T. Hubeny, 1996, \aaa, 309, 218,
\bibitem{} D.S. Leckrone, M.A.J. Snijders, 1979, \apjs, 39, 549,
\bibitem{} D.S. Leckrone, 1984, \apj, 286, 725,
\bibitem{} D.M. Pyper, 1969, \apjs, 18, 347,
\bibitem{} T. Ryabchikova, N. Piskunov, I. Savanov, F. Kupka, V. Malanushenko,
    1999, \aaa, 343, 229,
\bibitem{} T. Ryabchikova, 2008, Cont. Ast. Obs. Skalnat\'e Pleso, 38, 257,
\bibitem{} S.N. Shore, D.N. Brown, G. Sonneborn, 1987, \aj, 94, 737,
\bibitem{} N.A. Sokolov, 2000, \aaa, 353, 707,
\bibitem{} N.A. Sokolov, 2006, \mnras, 373, 666,
\bibitem{} N.A. Sokolov, 2010, \apss, 330, 37,
\bibitem{} N.A. Sokolov, 2011, this issue, p. 390
\bibitem{} K. Stepie{\' n}, W. Czechowski, 1993, \aaa, 268, 187,
\end{thebibliography}
\end{document}